\def\ispreprint{1}
\documentclass[journal]{IEEEtran}

\usepackage[utf8]{inputenc}
\usepackage[T1]{fontenc}
\usepackage{amsmath,amssymb}
\usepackage{graphicx}
\usepackage{circuitikz} 
\usetikzlibrary{positioning} 
\usepackage{siunitx}
\usepackage{booktabs}
\usepackage[hidelinks]{hyperref}
\usepackage{cite}

\graphicspath{{figures/}}
\newcommand{\iaopp}{IA-OPP}
\newif\ifrevmarks
\ifdefined\ismarked\revmarkstrue\fi
\newcommand{\rev}[1]{{\ifrevmarks\color{blue}\fi #1}}
\newcommand{\paperfig}[2][\columnwidth]{%
  \IfFileExists{figures/#2.pdf}%
    {\includegraphics[width=#1]{#2}}%
    {\fbox{\parbox[c][0.55\columnwidth][c]{0.95\columnwidth}%
      {\centering pending figure: \texttt{\detokenize{#2}.pdf}\\
       (see figures/FIGURES\_TODO.md)}}}}

\begin{document}

\title{Impedance-Aware Optimized Pulse Patterns for\\
Reconfigurable Battery Systems}

\author{Julian~Estaller,
        Johannes~Buberger,
        Andreas~Wiedenmann,
        Wolfgang~Grupp,\\
        Tobias~H{\"o}gerl,
        and~Thomas~Weyh
\thanks{The authors are with the Chair of Electrical Energy Supply,
University of the Bundeswehr Munich, 85577 Neubiberg, Germany
(e-mail: julian.estaller@unibw.de).}
\ifdefined\ispreprint
\thanks{This work has been submitted to the IEEE for possible
publication. Copyright may be transferred without notice, after which
this version may no longer be accessible.}%
\else
\thanks{Manuscript received ...; revised ....}%
\fi
}

\ifdefined\ispreprint
\markboth{Preprint --- submitted to the IEEE Open Journal of Power Electronics}%
{Estaller \MakeLowercase{\textit{et al.}}: Impedance-Aware Optimized Pulse Patterns for Reconfigurable Battery Systems}
\else
\markboth{IEEE Open Journal of Power Electronics (draft)}%
{Estaller \MakeLowercase{\textit{et al.}}: Impedance-Aware Optimized Pulse Patterns for Reconfigurable Battery Systems}
\fi

\maketitle

\begin{abstract}
Reconfigurable battery systems (RBS) synthesize the converter output
voltage by inserting and bypassing battery cell groups, so every switching
state changes not only the source voltage but also the source impedance.
This paper shows that this configuration-dependent impedance breaks a
tacit assumption of optimized pulse patterns (OPP): patterns designed
with the customary linear plant model overestimate their achievable
current quality by up to an order of magnitude and converge to a
distortion floor set by the resistance modulation. We propose
impedance-aware optimized pulse patterns (\iaopp{}), which embed the
level-dependent source impedance in the optimization objective, combined
with an event-budget formulation that treats the reconfiguration rate of
the communication bus as a managed resource. An exact closed-form
evaluation of the piecewise-exponential load current renders the
objective smooth in the switching instants. In a simulated
cascaded-bridge reference system, \iaopp{} attains 0.52\,\% current
total harmonic distortion (THD)
at a reconfiguration rate of 2.4\,kHz, outperforming nearest-level
modulation, phase-shifted carrier pulse-width modulation (PWM) at up to
five times the switching rate, and a classical OPP baseline at every event budget; the advantage
persists across the modulation-index range and under
$\boldsymbol{\pm}$30\,\% parameter mismatch. For three-wire
star-connected systems, masking triplen orders in the objective restores
the full advantage, with the impedance-aware patterns beating their
linear-model counterparts by about 20\,\% at equal budget.
\end{abstract}


\section{Introduction}
\label{sec:introduction}

\IEEEPARstart{R}{econfigurable} battery systems (RBS) move the power
electronics into the battery itself: half-bridge submodules (SMs) insert
or bypass individual cells or cell groups, so the converter synthesizes its
ac output voltage directly from the battery in cell-voltage granularity
\cite{goetz2015,goetz2017,chatzinikolaou2016}. In a cascaded-bridge
realization, the submodules of each module set the voltage
\emph{magnitude} in fine steps while an H-bridge per module provides the
polarity. For modulation this yields an unusually rich actuator: \rev{the
phase voltage is a staircase whose step \emph{instants} and whose
(integer) step \emph{heights} are both design variables, a freedom that
the fine cell-level granularity of the RBS makes meaningful, in contrast
to conventional multilevel converters whose few, coarse levels
effectively fix the staircase sequence.} The task of
the modulator is to convert this freedom into current quality without
losing sight of the cost of switching.

The application scenario considered here is a motor-type load,
represented by a series RL surrogate. It shapes the requirement profile
in a characteristic way. Operation is stationary and periodic, with the
reference constant over many fundamental periods. What matters is the
quality of the \emph{current}, since low-order current harmonics cause
additional losses and torque ripple\rev{, and degrade power quality
wherever converters couple to the grid \cite{adak2024}}. At the same
time, every switching
action is expensive: it causes losses in the semiconductors and, in a
reconfigurable system, occupies the
communication bus that coordinates the submodules. The goal is therefore
maximum current quality at a minimal, \emph{predictable} number of
switching events. This is the domain of optimized pulse
patterns (OPP) \cite{buja1977,holtz2007,rathore2010}: when the operating
point varies slowly, switching patterns can be computed offline to be
optimal over a full period, stored in a lookup table, and replayed
online. For an RBS feeding a motor-type load, OPPs are therefore a
natural choice.

On closer inspection, however, no existing member of the class fits the
system, because reconfigurable battery systems violate two basic
assumptions of the classical OPP formulation. The first concerns the
source. The configuration of an RBS determines its impedance: every
voltage level corresponds to a distinct configuration and hence a
distinct path resistance, so the source resistance is modulated
\emph{synchronously with the switching pattern}. Classical OPP
optimization, in contrast, presumes an ideal or at least
configuration-independent source. A pattern that is optimal for such a
model is no longer optimal for the real plant, and the deviation is
systematic. The same property is also an opportunity: a
method that knows the configuration-dependent impedance in its
objective can exploit it instead of optimizing past it.

The second peculiarity concerns the cost of switching. \rev{Classical
OPP formulations use the number of switching events per fundamental
period as the single switching-effort measure: selected per operating
point from the lookup table, but serving simultaneously as
switching-frequency and loss measure, which is adequate as long as every
event switches one step of topology-given height. In an RBS the two
roles decouple}: a level change is one string reconfiguration
and loads the communication bus exactly once, regardless of how many
submodules switch in the process, while the number of submodule
switching actions governs the switching losses. \rev{Together with
switching events that may span several levels at once,} this creates a
cost structure that a single pulse number no
longer describes. The appropriate formulation is an \emph{event budget}
of level changes per period, assessed by a dual metric of bus events and
submodule switching actions.

This leads to the thesis of the paper: OPP is the right class of
methods for reconfigurable battery systems, but only a formulation that
incorporates both structural properties, the configuration-dependent
source
impedance and the event-based cost structure, into the optimization
unlocks
the potential of the system. We develop such a formulation as
impedance-aware optimized pulse patterns (\iaopp{}). \rev{To the best of
the authors' knowledge, no existing modulation method embeds the
configuration-dependent source impedance in the pattern-shaping
objective itself; Section~\ref{sec:related} substantiates this against
the closest related work.} The contributions
are:
\begin{enumerate}
  \item an OPP objective that models the level-dependent source
    impedance of the RBS and thereby minimizes the current distortion of
    the \emph{real} plant (Sections~\ref{sec:model}
    and~\ref{sec:method});
  \item a cost formulation matched to the event-based structure of the
    system: an exact event budget per period with \rev{optimized
    multi-level step sizes},
    assessed by the dual metric of string reconfigurations and submodule
    switching actions, instead of a fixed pulse number as the sole
    switching measure (Section~\ref{sec:method});
  \item an exact, sampling-free closed-form evaluation of the switched
    RL plant under level-synchronously varying path resistance, which
    yields exact Fourier coefficients and loss figures and places the
    optimization and every comparison in this paper on an identical
    footing (Section~\ref{sec:model});
  \item a quantitative measurement of the value of plant modeling via a
    single-variable experiment against a classically formulated OPP
    baseline at identical budget, together with a benchmark against
    carrier-based and reference-tracking methods, a robustness analysis,
    and a three-phase extension in which a triplen-masked objective
    restores the full advantage (Sections~\ref{sec:results}
    and~\ref{sec:threephase}).
\end{enumerate}

The study is simulation-based throughout; all quantitative claims rest
on an exact analytical evaluation of the switched RL plant rather than
on time-stepping simulation, and experimental validation remains future
work (Section~\ref{sec:conclusion}). The remainder of the paper is
organized as follows. Section~\ref{sec:related} positions the method
within the OPP literature and the modulation literature on
reconfigurable batteries. Section~\ref{sec:model} introduces the
reference system, the level-synchronous resistance modulation, and the
closed-form evaluation. Section~\ref{sec:method} develops the \iaopp{}
optimization. Section~\ref{sec:results} reports the single-phase
results, including the classical-OPP baseline experiment, the benchmark,
and the robustness analysis. Section~\ref{sec:threephase} extends the
analysis to three-wire star-connected systems, and
Section~\ref{sec:conclusion} concludes.

\section{Related Work}
\label{sec:related}

The idea of optimizing the switching angles of a periodic waveform
offline against a spectral criterion goes back to the selective harmonic
elimination (SHE) of Patel and Hoft \cite{patel1973,patel1974}. It was
soon complemented by the minimization of a scalar distortion measure
\cite{buja1977} and by the synchronous optimal pulse-width modulation
(PWM) developed by Holtz
and coworkers for high-power drives, later extended to multilevel
converters \cite{holtz2007,rathore2010}. This line established the
conventions that the OPP family shares and that the present method
adopts without claiming them as contributions: quarter-wave symmetry to
reduce the search space, the number of switching events per period as
the design parameter, and the storage of patterns in offline lookup
tables over the modulation index. The state of the optimization art now
extends to certified lower bounds on the attainable distortion, with
modern formulations optimizing the sequence of voltage levels alongside
the switching instants \cite{miller2025}.

For multilevel converters, angle optimization was transferred to
staircase waveforms, first with a fixed level sequence
\cite{chiasson2004,dahidah2008,dahidah2015}, later for cascaded bridges
with unequal dc-link voltages \cite{tolbert2005} up to minimization of
the current total harmonic distortion (THD) with the source-voltage
ratios as additional design
variables \cite{barbie2023}\rev{, and to time-domain formulations that
minimize voltage and current THD of the staircase directly
\cite{bairamkulov2016}}. These works share one modeling assumption:
the sources may be unequal, but they are \emph{constant}. The
optimization evaluates the voltage spectrum, or the current of a linear
surrogate plant, while the source itself remains unaffected by the
switching state. A source
impedance that is \emph{modulated} synchronously with the levels, as
exhibited by a reconfigurable battery system, lies outside this model
class. The same holds for OPP formulations for modular multilevel
converters \cite{huber2012}, which account for internal system states
but not for a configuration-dependent source impedance.

For battery-integrated converters---the system class of this
paper---modulation has so far mainly served battery management: works on
nearest-level and carrier-based schemes use the reconfiguration freedom
for state-of-charge balancing, loss and temperature distribution, and
the mitigation of ripple-induced aging
\cite{goetz2015,goetz2017,kersten2019,altaf2017,kacetl2023,kacetl2022}\rev{;
related control work schedules the reconfiguration of battery strings at
fast-charging stations by rule-based and model-predictive policies
\cite{pinter2025}: online decision making above the modulation layer,
not pattern synthesis. Closest in spirit is the hybrid modulation of
Kersten \emph{et al.} \cite{kersten2021}, which models battery and
inverter losses of a battery-fed cascaded bridge and uses them to
\emph{select}, per operating point, between two standard schemes
(multilevel PWM and fundamental-frequency selective harmonic
elimination). There, the modeled impedance decides \emph{which}
conventional pattern is replayed; it does not shape the pattern itself,
whose switching angles remain classical impedance-free solutions}. That the
configuration simultaneously determines the impedance of the system is
known in this literature as a system property and is used, e.g., in loss
and ripple models. As part of a \emph{modulation objective}, i.e., as a
quantity that shapes the switching pattern itself, it has to the best
of the authors' knowledge not been used. The most recent work confirms
this picture: current research on modulation optimization for
reconfigurable batteries shifts carrier phases online by means of neural
networks against a THD criterion on a linear model
\cite{hashemizadeh2025}. That approach is carrier-based, greedy in
operation, and uses no impedance in its plant model; it is complementary
to the
offline, period-optimal, impedance-aware patterns pursued here. Finally,
finite-control-set model predictive control (FCS-MPC) does bring a
plant model into the switching
decision \cite{rodriguez2013}, but optimizes greedily over short
horizons at every sampling step: neither does a period-optimal pattern
emerge, nor is the number of events per period deterministic, a
structural disadvantage for the bus-bound operation of an RBS.

Table~\ref{tab:contrast} condenses this positioning. \iaopp{} differs
from the optimizing methods in three characteristics: the plant model
(level-dependent source impedance), the objective evaluated on it
(current THD of the real plant), and the cost formulation: an exact
event budget per period with optimized step sizes, assessed by the dual
metric of string reconfigurations and submodule switching actions,
instead of a single pulse number as the switching measure. The
freedom of the level sequence taken by itself is shared with recent OPP
formulations \cite{miller2025}; what is new is its coupling to the
event-based cost structure of the RBS. Conversely, we do not claim
optimized angle computation as such, quarter-wave symmetry, or the
lookup-table principle; in these respects the method
stands in the tradition of the OPP family and inherits its maturity and
operational track record. The quantitative evidence that the plant model
carries the difference follows in Section~\ref{sec:results} through the
direct comparison with a classically formulated OPP baseline at
identical budget.

\begin{table}[!t]
  \caption{Positioning of \iaopp{} against the optimizing modulation
    methods.}
  \label{tab:contrast}
  \centering
  \renewcommand{\arraystretch}{1.2}
  \setlength{\tabcolsep}{1.5pt}
  \footnotesize
  \begin{tabular}{lcccc}
    \toprule
    & SHE-PWM & OPP & FCS-MPC & \textbf{\iaopp{}} \\
    \midrule
    Design & offline & offline & online & \textbf{offline} \\
    Switching instants & optimized & optimized & grid-bound
      & \textbf{optimized} \\
    \rev{Levels per event} & \rev{single} & \rev{single$^{\mathrm{a}}$}
      & \rev{free}
      & \rev{\shortstack{\textbf{optimized}\\\textbf{multi-level}}} \\
    Objective & \shortstack{selective\\elimination} &
      \shortstack{THD-optimal\\(voltage)} &
      \shortstack{cost function\\per step} &
      \shortstack{\textbf{THD-optimal}\\\textbf{(current)}} \\
    Plant model & \shortstack{none\\(ideal source)} &
      \shortstack{ideal to\\linear} &
      \shortstack{linear\\(prediction)} &
      \shortstack{\textbf{level-dep.}\\\textbf{impedance}} \\
    \shortstack[l]{Events\\per period} &
      \rev{\shortstack{pulse no.\ per\\op.\ point}} &
      \rev{\shortstack{pulse no.\ per\\op.\ point}} & variable &
      \shortstack{\textbf{budgeted}\\\textbf{($\boldsymbol{n}$)}} \\
    \bottomrule
  \end{tabular}
  \par\vspace{2pt}
  {\footnotesize \rev{$^{\mathrm{a}}$\,the level \emph{sequence} is
  co-optimized in recent formulations \cite{miller2025}.}}
\end{table}

\section{System Model and Evaluation Metrics}
\label{sec:model}

\subsection{Reference System and Operating Point}
\label{ssec:refsystem}

\begin{figure}[!t]
  \centering
    \begin{circuitikz}[american, circuitikz/bipoles/length=0.9cm,
      every node/.style={font=\scriptsize}, scale=0.88, transform shape]
    \node at (-0.15,2.05) {\textbf{(a)}};
    \node[draw, minimum width=1.0cm, minimum height=0.5cm] (m1) at (1.0,1.3) {Mod.\,1};
    \node[draw, minimum width=1.0cm, minimum height=0.5cm] (m2) at (2.35,1.3) {Mod.\,2};
    \node at (3.3,1.3) {$\cdots$};
    \node[draw, minimum width=1.0cm, minimum height=0.5cm] (m5) at (4.25,1.3) {Mod.\,5};
    \draw (0,1.3) to[short, i=$i$] (m1.west);
    \draw (m1.east) -- (m2.west);
    \draw (m2.east) -- (3.05,1.3);
    \draw (3.55,1.3) -- (m5.west);
    \draw (m5.east) -- (5.9,1.3) -- (5.9,-0.3) -- (4.7,-0.3);
    \draw (4.7,-0.3) to[R=$R$] (3.0,-0.3) to[L=$L$] (1.3,-0.3)
      -- (0,-0.3) -- (0,1.3);
    \draw[dashed, rounded corners=2pt] (1.1,-0.75) rectangle (4.9,0.3);
    \node[anchor=west] at (1.2,-0.55) {load};
    \draw[->] (4.9,0.75) -- node[below]{$v_\mathrm{s}$} (0.35,0.75);
    \begin{scope}[yshift=-5.15cm]
      \node at (-0.15,3.35) {\textbf{(b)}};
      \node[draw, minimum width=0.9cm, minimum height=0.42cm] (sm1) at (0.75,2.6) {SM\,1};
      \node[draw, minimum width=0.9cm, minimum height=0.42cm] (sm2) at (0.75,1.95) {SM\,2};
      \node[draw, minimum width=0.9cm, minimum height=0.42cm] (sm12) at (0.75,0.85) {SM\,12};
      \fill (0.75,1.36) circle (0.6pt);
      \fill (0.75,1.47) circle (0.6pt);
      \fill (0.75,1.58) circle (0.6pt);
      \draw (sm1.south) -- (sm2.north);
      \draw (sm2.south) -- (0.75,1.68);
      \draw (0.75,1.26) -- (sm12.north);
      \draw (sm1.north) -- (0.75,3.05) -- (3.4,3.05);
      \draw (sm12.south) -- (0.75,0.4) -- (3.4,0.4);
      \draw (2.4,3.05) to[nos] (2.4,1.725);
      \draw (2.4,1.725) to[nos] (2.4,0.4);
      \draw (3.4,3.05) to[nos] (3.4,1.725);
      \draw (3.4,1.725) to[nos] (3.4,0.4);
      \draw (2.4,1.725) -- (2.9,1.725) -- (2.9,2.075) -- (3.26,2.075);
      \draw (3.54,2.075) -- (4.15,2.075) node[ocirc]{};
      \draw (3.4,1.725) -- (4.15,1.725) node[ocirc]{};
      \node[anchor=west] at (1.35,3.35) {H-bridge (polarity)};
      \node[rotate=90, anchor=south] at (0.1,1.75) {SM chain};
    \end{scope}
    \begin{scope}[yshift=-5.15cm, xshift=4.9cm]
      \node at (0.05,3.35) {\textbf{(c)}};
      \draw (0.4,2.5) to[battery1, l=$u_\mathrm{cell}$] (0.4,1.1);
      \draw (0.4,2.5) -- (0.4,3.05) -- (1.7,3.05);
      \draw (0.4,1.1) -- (0.4,0.4) -- (1.7,0.4);
      \draw (1.7,3.05) to[nos] (1.7,1.725);
      \draw (1.7,1.725) to[nos] (1.7,0.4);
      \draw (1.7,1.725) -- (2.5,1.725) node[ocirc]{};
      \draw (1.7,0.4) -- (2.45,0.4) node[ocirc]{};
      \node[anchor=west] at (0.75,3.35) {half-bridge};
    \end{scope}
  \end{circuitikz}
  \caption{Reference system. (a)~Single-phase string of five cascaded
    modules feeding the RL load. (b)~Module: twelve series submodules
    form the dc side of an H-bridge polarity stage. (c)~Submodule: a
    half-bridge inserts or bypasses one battery cell.}
  \label{fig:topology}
\end{figure}
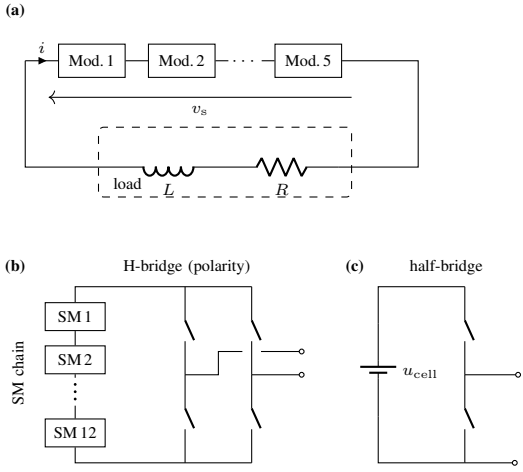

The reference system is the single-phase string of a cascaded-bridge
RBS (Fig.~\ref{fig:topology}): five series-connected modules, each
comprising twelve half-bridge
submodules with one battery cell each (cell voltage \SI{3.7}{\volt}).
The submodule layer thus provides up to 60 voltage levels per string
with a range of \SI{222}{\volt}; the H-bridges set the polarity. The
operating point represents a motor-type load at partial load by an RL
surrogate: fundamental frequency \SI{50}{\hertz}, voltage amplitude
\SI{80}{\volt} (a modulation index $m_\mathrm{a} \approx 0.36$, defined
as the ratio of the reference amplitude to the \SI{222}{\volt} maximum
string voltage), load resistance $R = \SI{2}{\ohm}$, and load inductance
$L = \SI{10}{\milli\henry}$.
No back electromotive force (back-EMF) is modeled; the surrogate
captures the stationary current
response only. Table~\ref{tab:parameters} summarizes the parameters.

\begin{table}[!t]
  \caption{Parameters of the reference system and the nominal
    operating point.}
  \label{tab:parameters}
  \centering
  \renewcommand{\arraystretch}{1.15}
  \setlength{\tabcolsep}{8pt}
  \small
  \begin{tabular}{llr}
    \toprule
    & Quantity & Value \\
    \midrule
    System & Modules per string & 5 \\
    & Submodules per module & 12 \\
    & Cell voltage & \SI{3.7}{\volt} \\
    & Number of levels (max.) & 60 \\
    & Cell internal resistance $r_\mathrm{bat}$ & \SI{10}{\milli\ohm} \\
    & Submodule on-resistance $r_\mathrm{SM}$ & \SI{1}{\milli\ohm} \\
    & H-bridge on-resistance $r_\mathrm{HB}$ & \SI{2}{\milli\ohm} \\
    \midrule
    Operating point & Fundamental frequency & \SI{50}{\hertz} \\
    & Voltage amplitude (reference) & \SI{80}{\volt} \\
    & Modulation index $m_\mathrm{a}$ & $\approx 0.36$ \\
    & Load resistance $R$ & \SI{2}{\ohm} \\
    & Load inductance $L$ & \SI{10}{\milli\henry} \\
    \bottomrule
  \end{tabular}
\end{table}

\subsection{Level-Synchronous Resistance Modulation}
\label{ssec:resmod}

In the reference system, the coupling between configuration and
impedance takes the following form. The effective path resistance at
voltage level
$k$ is
\begin{equation}
  R_\mathrm{path}(k) = R
    + 5 \cdot 2\,r_\mathrm{HB}
    + \left\lceil \tfrac{|k|}{12} \right\rceil \cdot 12\,r_\mathrm{SM}
    + |k| \cdot r_\mathrm{bat},
  \label{eq:pathres}
\end{equation}
with the load resistance $R$, two conducting semiconductors per H-bridge
($r_\mathrm{HB} = \SI{2}{\milli\ohm}$, in the path even when a module is
bypassed), the submodule contribution
($r_\mathrm{SM} = \SI{1}{\milli\ohm}$), and the internal resistances of
the inserted cells ($r_\mathrm{bat} = \SI{10}{\milli\ohm}$). The
submodule term changes module-wise: in every active module the entire
submodule chain carries the current (one conducting semiconductor per
submodule, whether its cell is inserted or bypassed); bypassed
modules contribute nothing. Only the battery term grows
linearly with the level. At the nominal operating point (levels up to
$\pm 22$ in use), the path resistance therefore spans the range from
\SI{2.020}{\ohm} (zero level) to \SI{2.264}{\ohm}, a level-synchronous
modulation of \SI{12}{\percent} relative to the zero level. \rev{Which
levels are in use, and hence the span, is itself an operating-point
property: the modulation index sets the required peak level, so
\eqref{eq:pathres} maps every operating point and every pattern to its
own resistance trajectory; all figures quoted in this subsection refer
to the nominal point.} The plant is
therefore not time-invariant: it is modulated in step with the
switching pattern. Section~\ref{ssec:plantvalue} quantifies the
consequences for methods designed on a time-invariant model.

\rev{Since the level \emph{magnitude} $|k(t)|$ repeats with half the
fundamental period, the path resistance along a given pattern is an
even-harmonic Fourier series,
\begin{equation}
  R_\mathrm{path}(t) = \bar{R}
    + \sum_{\nu \ge 1} \hat{R}_{2\nu} \cos(2\nu\omega t + \varphi_{2\nu}),
  \label{eq:resfourier}
\end{equation}
whose coefficients follow from \eqref{eq:pathres} and the pattern. For a
voltage-faithful staircase at the nominal point, $\bar{R} =
\SI{2.18}{\ohm}$, dominated by $\hat{R}_2 = \SI{99.5}{\milli\ohm}$, a
modulation depth of \SI{4.6}{\percent}. In the voltage balance
$v(t) = R_\mathrm{path}(t)\, i(t) + L\, \mathrm{d}i/\mathrm{d}t$, the
product of the even resistance series \eqref{eq:resfourier} with the odd
current spectrum generates exclusively \emph{odd} current harmonics; the
dominant term mixes $\hat{R}_2$ with the fundamental $\hat{I}_1$ onto
the third order, where the plant impedance
$Z(3\omega) = \bar{R} + \mathrm{j}\,3\omega L$ attenuates it to
\begin{equation}
  \hat{I}_3 \approx \frac{\hat{R}_2\, \hat{I}_1}{2\, |Z(3\omega)|}
    = \SI{0.107}{\ampere}.
  \label{eq:mixing}
\end{equation}
The exact evaluation of the same staircase yields \SI{0.114}{\ampere};
the remaining \SI{7}{\percent} stem from higher-order mixing cascades.}
This mechanism will reappear throughout
the paper as a distortion floor for all voltage-oriented methods.

\subsection{Closed-Form Evaluation of the Switched RL Plant}
\label{ssec:closedform}

Between two switching events, the plant sees the constant staircase
source voltage $u_n$ at \rev{the constant path resistance
$R_n = R_\mathrm{path}(k_n)$ of the active level $k_n$}. The current
follows the
closed-form solution
\begin{equation}
  i(t) = \frac{u_n}{R_n}
    + \left( i(t_n) - \frac{u_n}{R_n} \right)
      \mathrm{e}^{-(t - t_n)/\tau_n},
  \qquad \tau_n = \frac{L}{R_n},
  \label{eq:segment}
\end{equation}
for $t \in [t_n, t_{n+1})$, where the segment initial values $i(t_n)$
are chained by continuity and
the periodicity of the steady-state solution closes the system of
equations. \rev{Note that the level-dependent path resistance enters
both the stationary gain $u_n/R_n$ and the time constant $\tau_n$ of
every segment: the source impedance shapes the current \emph{dynamics}
segment by segment, not merely a dc operating point. (What
\eqref{eq:segment} does idealize is the cell impedance itself, taken as
purely ohmic---see the limitations in Section~\ref{sec:conclusion}.)} Since \eqref{eq:segment} consists piecewise of exponential
and constant terms, all evaluation quantities---the Fourier coefficients
of the current, root-mean-square (RMS) values, active and apparent
power, and the ohmic
losses---can be integrated segment-wise \emph{exactly}. Time-stepping
simulation is dispensed with entirely. This has two
consequences. First, all metrics are free of sampling and
discretization errors and smooth in the switching instants, the
prerequisite for the continuous timing refinement in
Section~\ref{sec:method}. Second, every method in this
paper---\iaopp{}, baselines, and reference methods---passes through the
same closed-form evaluation, so differences in the results are
attributable exclusively to the patterns themselves.

\rev{To rule out implementation errors in the closed-form chain, the
evaluator was additionally cross-checked against an independent
time-stepping simulation of the switched plant: a fourth-order
Runge--Kutta integration of the piecewise differential equation
(step size $\le \SI{0.5}{\micro\second}$, aligned to the switching
edges), implemented without shared code, settling into steady state
from zero initial current instead of using the analytical periodicity
condition. For the selected \iaopp{} pattern, 1S-NLM, and
submodule-level PS-PWM, the two implementations agree to within
$10^{-7}$ percentage points of THD and $3\times10^{-13}$~A of
instantaneous current, i.e., to double-precision limits.}

\subsection{Evaluation Metrics}
\label{ssec:metrics}

Current quality is assessed as the THD of the load current,
\begin{equation}
  \mathrm{THD}_i = \frac{\sqrt{\sum_{h=2}^{50} \hat{I}_h^2}}{\hat{I}_1},
  \label{eq:thd}
\end{equation}
with the exactly computed harmonics $\hat{I}_h$ up to order 50.
Restricting the band is uncritical in the reference system because the
load inductance strongly attenuates higher-frequency components;
in-band and wide-band THD practically coincide. As the feasibility
condition, the fundamental amplitude of the staircase \emph{source}
voltage (computed exactly from its Fourier series) must match the
\SI{80}{\volt} reference within \SI{1}{\percent}; the current and
load-voltage fundamentals do not enter feasibility. The phase is fixed
by construction for quarter-wave-symmetric patterns and poses no
separate condition. Losses follow in closed form from the segment-wise
$R_n$-weighted integrals of the squared current.

The cost of switching is captured by the dual metric motivated in
Section~\ref{sec:introduction}. The number of level changes per
fundamental period, $n$, counts the string reconfigurations and is the
measure of bus load; at \SI{50}{\hertz}, $n = 48$ corresponds to a bus
rate of \SI{2.4}{\kilo\hertz}. The number of submodule switching actions
per period captures the loss effort: a level change spanning several
levels is \emph{one} bus event but switches correspondingly many
submodules. In an RBS the two quantities decouple; only
their pair describes the switching cost completely.

\section{Impedance-Aware Optimized Pulse Patterns}
\label{sec:method}

\subsection{Pattern Structure and Search Space}
\label{ssec:structure}

\iaopp{} optimizes staircase patterns under three structural
constraints. First, the pattern is \emph{quarter-wave symmetric}: all
even harmonics and all cosine components vanish, and the search space
reduces to the first quarter period, the classical OPP convention
\cite{patel1973}. Second, the quarter wave is \emph{strictly
monotonically increasing}: every switching edge raises the level, and no
event is spent on up-and-down movement within the quarter wave. Third,
\rev{the \emph{step sizes are design variables}: an edge may switch
several submodule levels at once, the step height being an integer
multiple of the cell voltage chosen by the optimization}. A pattern is thus completely described by the $K$ edge
instants and the associated level heights of the first quarter period;
per fundamental period, $n = 4K$ level changes result. The event budget
$n$ (the bus load of Section~\ref{ssec:metrics}) is
the central design parameter; monotonicity ensures that every
budgeted event produces voltage swing.

\subsection{Two-Stage Optimization}
\label{ssec:twostage}

For each budget, the pattern is determined in two stages
\rev{(Fig.~\ref{fig:flow})}: a dynamic
program (DP) provides the level sequence and an initial placement of the
edge instants; a coordinate descent subsequently refines the edge
instants continuously on the exact plant model.

\begin{figure}[!t]
  \centering
  \rev{%
  \begin{tikzpicture}[font=\scriptsize, >=latex, node distance=3.5mm,
      box/.style={draw, align=center, minimum width=6.6cm,
        minimum height=5.5mm, inner sep=2pt}]
    \node[box] (op) {operating point: $m_\mathrm{a}$, $f_0$, $R$, $L$,
      $R_\mathrm{path}(k)$ from \eqref{eq:pathres}};
    \node[box, below=of op] (dp) {budget DP on adaptive grid
      \eqref{eq:vreq}:\\ level sequence + initial edge instants,
      one candidate per budget $n$};
    \node[box, below=of dp] (ref) {coordinate-descent timing refinement
      \eqref{eq:objective}\\ on the exact closed-form plant
      \eqref{eq:segment}};
    \node[box, below=of ref] (sel) {selection rule: feasible $\to$
      min.\ THD $\to$ smaller budget\\ (or target-THD mode)};
    \node[box, below=of sel] (lut) {offline lookup table $\to$ online
      replay};
    \draw[->] (op) -- (dp);
    \draw[->] (dp) -- (ref);
    \draw[->] (ref) -- (sel);
    \draw[->] (sel) -- (lut);
  \end{tikzpicture}}
  \caption{\rev{Overview of the \iaopp{} pattern synthesis: offline
    two-stage optimization per operating point, deterministic selection,
    lookup-table replay online.}}
  \label{fig:flow}
\end{figure}

The DP operates on a discretization of the quarter period into 90
intervals. The grid is adaptive: the interval density follows
$0.30 + 0.70\,|\cos(\omega t)|$, dense at the zero crossing of the
reference, where it is steep and edge instants strongly influence the
result, and coarser at the crest. The state space comprises the interval
index, the voltage level, and the number of level changes already spent;
admissible transitions are remaining on the level or switching to a
strictly higher one; the monotonicity of the quarter wave is thus built
directly into the transition structure. As additive cost, each interval
contributes the time-weighted squared error of the level voltage against
an \emph{impedance-compensated} reference,
\begin{equation}
  V_\mathrm{req}(t) = v_\mathrm{ideal}(t)
    + R_\mathrm{conv}(\ell)\, i_\mathrm{ideal}(t),
  \label{eq:vreq}
\end{equation}
i.e., the ideal load voltage plus the voltage drop that the
level-dependent converter-internal resistance share
$R_\mathrm{conv}(\ell)$ from \eqref{eq:pathres} causes at the ideal
current. Impedance awareness is thus anchored already in the
initialization: even the discrete candidate aims not at the ideal
voltage shape but at the predistorted shape that produces the ideal
current on the real plant. The DP criterion is a
surrogate that renders the problem additive and hence exactly solvable;
the refinement and the selection rule assess the actual current THD. For every reachable number of changes, the best terminal state is
reconstructed into a quarter-wave pattern, mirrored to the full period,
and evaluated exactly: one candidate per exact budget. The budget
is met exactly rather than as an upper bound; this yields the clean
mapping from bus load to attainable current
quality that Section~\ref{ssec:benchmark} evaluates.

\rev{This hybrid representation is deliberate, and it is why the
classical switching-angle parameterization is not used here: the angle
representation presumes a \emph{fixed} level sequence, which turns
pattern design into a purely continuous problem over the angles
\cite{patel1973,rathore2010}. With the level sequence and the step sizes
as design variables, the problem acquires a combinatorial layer that
angles cannot express. The DP solves exactly this layer, and it is not
an exhaustive search: with 90 intervals, 61 levels, and up to
$K = 24$ changes per quarter wave, the state space holds roughly
$1.3\times10^{5}$ states with ordered transitions, solved exactly in a
fraction of a second (Section~\ref{ssec:selection}). The subsequent
coordinate descent \emph{is} the continuous angle optimization of the
classical approach, executed on the exact plant model with the edge
instants as the angles.}

The second stage refines the edge instants at fixed level sequence by
cyclic coordinate descent: per edge, a one-dimensional minimization
(golden-section search with parabolic interpolation) is executed between
the neighboring edges, with a minimum spacing of
$T/7200 \approx \SI{2.8}{\micro\second}$ and a time resolution of
\SI{100}{\nano\second}; a new edge instant is accepted only upon strict
improvement. The passes over all edges terminate as soon as a full pass
improves the objective by less than $10^{-3}$ percentage points. This
direct search on the continuous parameter is possible because the
closed-form evaluation of Section~\ref{ssec:closedform} rates every
candidate configuration exactly and smoothly in the edge instants: the
optimizer is not fighting discretization noise. The refinement objective
is
\begin{equation}
  J = \mathrm{THD}_i + w \cdot \varepsilon_1 + B(\varepsilon_1),
  \qquad w = 0.01,
  \label{eq:objective}
\end{equation}
with the fundamental error $\varepsilon_1$ of the staircase source
voltage and a hard barrier $B$ upon violation of the \SI{1}{\percent}
tolerance. The fundamental error is thus not forced to zero but only
weakly penalized: within the tolerance band, the optimization may shift
the fundamental amplitude if doing so lowers the current THD. The choice
of $w$ is uncritical: the budget selection is invariant over more than
three orders of magnitude of $w$, and the attained THD lies on a plateau
(\SIrange{0.519}{0.521}{\percent}) for $w \le 1$; only a dominant
fundamental weight degrades the result markedly, because it subjects the
edge placement to the wrong goal. Decisive is that
\eqref{eq:objective} contains the current THD of the \emph{real} plant:
through the closed-form evaluation, the level-dependent path resistance
enters every evaluation. Together with \eqref{eq:vreq}, this is where
\iaopp{} departs from the classical formulation.

The refinement is a local search from a single DP initialization; a
multi-start probe at the selected budget shows this to be benign. Ten
runs at $n = 48$ with the initial edge instants jittered by up to
\SI{20}{\percent} of the local edge spacing all remain feasible and end
within \SIrange{0.513}{0.522}{\percent} THD (standard deviation 0.004
percentage points), with individual starts finding marginally better
local optima than the reported solution; the reported values are thus
conservative. Only at very small budgets ($n \le 20$) does the
single-start scheme scatter noticeably; the operating range of this
paper ($n \ge 24$) is unaffected.

\subsection{Selection Rule and Operating Modes}
\label{ssec:selection}

From the refined candidates, one per budget, a deterministic rule
selects: feasible is whoever meets the fundamental tolerance; among the
feasible, the minimum current THD wins; if candidates lie closer than
0.01 percentage points, the smaller budget wins. The tie-break encodes
the dual metric into the selection: current quality that differs only in
the fourth decimal place does not justify a higher bus load. Besides
selection at a prescribed budget, the method supports a target-THD mode:
the budgets are traversed in ascending order and the search stops at the
first budget meeting the required current quality. This yields the
minimum
bus load for a demanded quality, the question that arises in
applications.

Like classical OPP, the method is an offline approach: the patterns are
precomputed per operating point and stored as a lookup table; online,
only the replay of the switching table remains. The computational effort
of the optimization is incurred once, outside of operation, and it is
modest: a full budget sweep at one operating point ($n = 4$ to $96$,
i.e., 24 candidate optimizations) completes in \SI{5.8}{\second} in a
single-threaded MATLAB implementation on a current desktop-class CPU,
of which the dynamic program and candidate construction take
\SI{0.2}{\second} and the timing refinement the remainder. Populating a
lookup table over the modulation-index range is thus a matter of
minutes.

\section{Simulation Results}
\label{sec:results}

All following results refer to the nominal operating point of
Table~\ref{tab:parameters}; every method passes through the identical
closed-form evaluation of Section~\ref{ssec:closedform}. The reference
methods are nearest-level modulation with fixed step size $S$ ($S$-NLM,
where 1S-NLM switches single submodule levels) and phase-shifted carrier
PWM (PS-PWM) with carrier-to-fundamental frequency ratio $m_f$, applied
at the submodule level and at the module level; the
classical OPP baseline is introduced in Section~\ref{ssec:plantvalue}.

\subsection{Selected Solution}
\label{ssec:chosen}

\begin{figure}[!t]
  \centering
  \paperfig{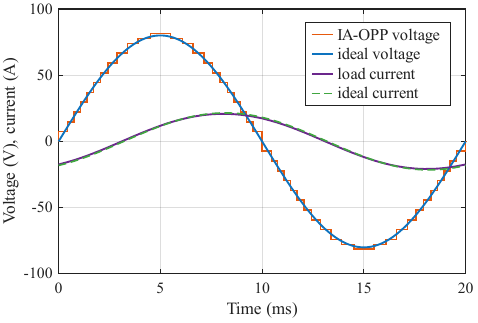}
  \caption{Selected \iaopp{} solution ($n = 48$) at the nominal
    operating point: staircase source voltage and load current over one
    fundamental period, each with its ideal reference \rev{(``ideal'':
    a lossless converter driving the pure RL load)}.}
  \label{fig:chosen}
\end{figure}

With free budget selection, the selection rule picks the pattern with
$n = 48$ level changes per period: current THD \SI{0.521}{\percent},
source-voltage fundamental amplitude \SI{80.507}{\volt} (error
\SI{0.633}{\percent}, within tolerance), and losses \SI{34.8}{\watt}. In
the dual metric this corresponds to a bus load of \SI{2.4}{\kilo\hertz}
at 88 submodule switching actions per period (\SI{4.4}{\kilo\hertz}).
Fig.~\ref{fig:chosen} shows the pattern over one fundamental period: the
staircase voltage hugs the ideal voltage with a visibly nonuniform edge
distribution, dense near the zero crossings, where the reference is
steep, with wide plateaus at the crest. The resulting load current is
barely distinguishable from the ideal current at this scale.

The current-error waveforms of the reference
methods at comparable switching effort separate into two components.
The ideal current
is defined against the pure load ($R$ and $L$, without
converter impedance): the output of a lossless converter.
Accordingly, all methods share a slowly varying error component of
nearly identical shape: the fundamental response to the time-weighted
mean series resistance of the converter of about \SI{0.18}{\ohm}, which
lowers the current fundamental from \SI{21.5}{\ampere} to
\SI{21.1}{\ampere} and shifts its phase by roughly \ang{2}. This
component is common to all methods and neutral for the THD, which counts
only orders $h \ge 2$. For \iaopp{} this component is slightly
smaller than the mean impedance suggests
(\SI{0.83}{\ampere} instead of \SI{0.93}{\ampere} peak), a first
effect of the predistortion isolated in
Section~\ref{ssec:plantvalue}. The methods differ in
the superimposed switching ripple, the quantity the THD measures. A
coarse 7S staircase buys its low event count with wide error arcs
(\SI{1.09}{\percent}\rev{; included for illustration only, since fixed
steps $S \ge 4$
cannot meet the fundamental tolerance at this operating point, cf.\
Section~\ref{ssec:benchmark}}). Module-level PS-PWM imprints hard
ripple spikes
(\SI{0.75}{\percent}). Submodule-level PS-PWM (\SI{0.57}{\percent}) and
2S-NLM (\SI{0.56}{\percent}) lie close together. \iaopp{} shows the
smoothest trace (\SI{0.52}{\percent})\rev{, at a bus load undercut
only by the two fixed-step staircases: marginally by the 2S
(\SI{2.2}{\kilo\hertz} versus \SI{2.4}{\kilo\hertz}, at 0.04
percentage points higher THD) and, at the price of infeasibility, by
the 7S}.

Table~\ref{tab:dualmetric} collects the dual metric, the current THD,
and the conduction losses of the main candidates. Two observations.
First, the table makes the decoupling of the two switching metrics
concrete: for PS-PWM every switching action is a string reconfiguration
(the two rates coincide), whereas the staircase methods separate
them: \iaopp{} pairs \SI{2.4}{\kilo\hertz} of bus load with
\SI{4.4}{\kilo\hertz} of submodule actions, and module-level PS-PWM
turns a moderate bus rate into a thirteenfold submodule rate because
every module event switches twelve submodules. Second, the conduction
losses lie in a narrow band (\SIrange{32.9}{34.8}{\watt}, or 7.8 to
\SI{8.0}{\percent} of the delivered fundamental power\rev{; for scale,
the \SI{2}{\ohm} load itself dissipates the delivered
$\approx\SI{437}{\watt}$; the table reports the converter-internal
share only}); the absolute
differences stem mainly from the delivered power itself: \iaopp{},
exploiting the tolerance band with a \SI{+0.6}{\percent} fundamental
amplitude, also delivers the most power. The loss figures are
conduction losses only: switching losses are not modeled, and precisely
for them the submodule switching rate remains the proxy: module-level
PS-PWM shows the lowest conduction losses in the field, but at a
submodule rate that real switching losses would dominate.
\rev{A datasheet-based estimate for the 3.7-V submodule MOSFETs
(switching energy of roughly \SI{1.3}{\micro\joule} per action at the
mean event currents of the patterns) puts the switching losses between
\SI{5.7}{\milli\watt} for \iaopp{} and \SI{71}{\milli\watt} for
module-level PS-PWM at its thirteenfold submodule rate, two to three
orders of magnitude below the conduction losses. At cell-voltage level,
conduction dominates the semiconductor losses; the submodule rate
remains the proxy that conservatively bounds the switching share.}

\begin{table}[!t]
  \caption{Dual switching metric, current THD, and conduction losses of
    the main candidates at the nominal operating point.}
  \label{tab:dualmetric}
  \centering
  \renewcommand{\arraystretch}{1.15}
  \setlength{\tabcolsep}{3pt}
  \footnotesize
  \begin{tabular}{lcccc}
    \toprule
    Method & \shortstack{Bus rate\\(kHz)} & \shortstack{SM rate\\(kHz)}
      & \shortstack{THD\\(\%)} & \rev{\shortstack{Conv.\ cond.\\losses (W)}} \\
    \midrule
    \iaopp{} ($n=48$)          & 2.4  & 4.4  & 0.521 & 34.8 \\
    Classical OPP ($n=48$)     & 2.4  & 4.2  & 0.607 & 33.2 \\
    1S-NLM                     & 4.4  & 4.4  & 0.561 & 34.2 \\
    2S-NLM                     & 2.2  & 4.4  & 0.562 & 34.5 \\
    PS-PWM submodule ($m_f=2$) & 11.8 & 11.8 & 0.571 & 34.1 \\
    PS-PWM module ($m_f=10$)   & 4.8  & 57.6 & 0.750 & 32.9 \\
    \bottomrule
  \end{tabular}
\end{table}

\subsection{Spectral Distribution}
\label{ssec:spectrum}

\begin{figure}[!t]
  \centering
  \paperfig{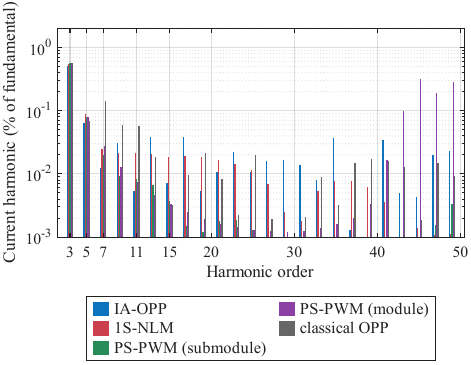}
  \caption{Current harmonics at the nominal operating point
    (logarithmic, relative to the fundamental) for \iaopp{}, 1S-NLM,
    PS-PWM at submodule and module level, and the classical OPP baseline
    at identical budget.}
  \label{fig:spectrum}
\end{figure}

Fig.~\ref{fig:spectrum} shows the current harmonics up to order 50.
Three observations follow. First, all methods share a
dominant third harmonic of about \SI{0.5}{\percent}; it alone accounts
for the bulk of the THD of every candidate, and
Section~\ref{ssec:plantvalue} isolates the underlying
method-independent mechanism. Second, the methods differ in where they
place the remaining distortion: 1S-NLM spreads it broadband over the odd
orders, whereas \iaopp{} suppresses the low orders and shifts
residual energy toward medium orders, where the load inductance
attenuates it more strongly. Submodule-level PS-PWM keeps the considered
band almost
empty but pays with carrier energy above the band and high switching
effort: 2.7 times the submodule switching rate and, since each of its
switching actions is also a string reconfiguration, 4.9 times the bus
load of the \iaopp{} solution. For module-level PS-PWM the carrier
sidebands already grow into the band from $h \approx 43$. Third, the
classical OPP baseline carried along (identical budget $n = 48$) shows
elevated harmonics at $h = 7$ to $11$, the
effect of misplaced edges that the next subsection quantifies.

\subsection{Value of Plant Modeling}
\label{ssec:plantvalue}

The central claim, that only impedance in the objective unlocks the
potential of the system, can be measured directly. To this end, a
classical OPP baseline is constructed that differs from \iaopp{} in
exactly one respect: the same optimizer with the same pattern structure,
the same budgets, the same refinement, and the same selection rule rates
its candidates during optimization on a \emph{linear} plant model. The
converter-internal resistances are set to zero in the objective, exactly
as in the classical OPP criterion of Section~\ref{sec:related}. With the
converter resistances zeroed, the current THD of the linear RL plant is
equivalent to an impedance-weighted voltage distortion, the customary
weighted-THD (WTHD) criterion of Table~\ref{tab:contrast};
evaluating the
baseline by current THD rather than voltage THD makes it, if anything,
the \emph{stronger} variant of the classical formulation.
\rev{Formally, the baseline runs the identical pipeline of
Fig.~\ref{fig:flow} with $R_\mathrm{conv}(\ell) \equiv 0$ in
\eqref{eq:vreq} and with the THD in \eqref{eq:objective} evaluated on
\eqref{eq:segment} at $R_n \equiv R$: the current-THD-optimal staircase
for an ideal source driving the RL load, the criterion of the
classical current-THD line \cite{buja1977,bairamkulov2016}.} The
finished patterns of both methods then pass through the same exact
evaluation on the real plant. Any difference in the outcome is thus
causally attributable to the plant model.

\subsubsection{Prediction Gap of the Linear Model}

The first finding concerns the self-assessment of the classical
formulation: for budgets $n \ge 24$, the linear model promises current
THD values of \SIrange{0.047}{0.25}{\percent}; on the real plant the
same patterns attain \SIrange{0.56}{0.64}{\percent}, a deviation of up
to a factor of 10. Classical OPP solves its optimization problem
precisely, but the problem is the wrong one: the prediction promises a
current quality that
no switching budget, however large, can deliver.

\subsubsection{Resistance Modulation as a THD Floor}

The second finding explains \emph{why} the prediction fails. With
growing budget, the real THD of classical OPP converges to about
\SI{0.565}{\percent}, the same value at which 1S-NLM and
submodule-level PS-PWM saturate (Fig.~\ref{fig:thdplane}). This common
value is the floor set by the mechanism
of Section~\ref{ssec:resmod}: all these methods are voltage-oriented.
They reproduce the ideal voltage shape ever more faithfully, while the
plant translates voltage into current through the
modulated path resistance \eqref{eq:pathres}, whose dominant second
harmonic mixes with the fundamental onto the third current harmonic
\eqref{eq:mixing}. For
voltage-oriented methods this distortion is a floor that no amount of
switching effort breaks, because it does not originate in the pattern
but in the plant. (Since the third harmonic is a common-mode quantity in
a three-phase arrangement, the floor question poses itself anew there;
Section~\ref{sec:threephase} pursues it.)

A counter-experiment confirms this: if the same patterns are evaluated
on a plant with constant resistance, the floor vanishes. Submodule
PS-PWM drops to \SI{0.0002}{\percent}, 1S-NLM to \SI{0.073}{\percent}.
The floor is thus a property of the plant, not of the patterns. The
reverse direction is equally instructive: the \iaopp{} solution attains
only \SI{0.245}{\percent} on the constant-resistance plant, worse than
the PWM there, because its predistortion now corrects an error that no
longer exists. A pattern is optimal
only for the plant it was designed for.

\subsubsection{Implicit Predistortion}

\begin{figure}[!t]
  \centering
  \paperfig{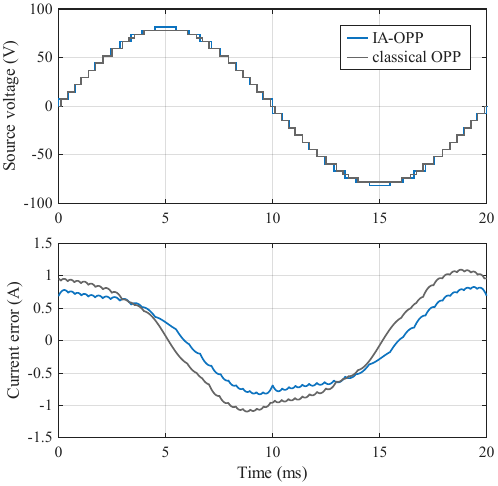}
  \caption{Implicit predistortion: \iaopp{} and classical OPP at
    identical budget $n = 48$: nearly congruent voltage staircases
    (top), clearly different current errors (bottom).}
  \label{fig:predistortion}
\end{figure}

How \iaopp{} undercuts the floor is shown in
Fig.~\ref{fig:predistortion} by the direct comparison of both methods at
identical budget $n = 48$. The voltage staircases (top) are nearly
congruent: the differences are limited to slightly shifted edges and a
marginally higher crest plateau. The current errors (bottom) differ
clearly: the error of classical OPP oscillates at about
$\pm$\SI{1.1}{\ampere} and carries the elevated harmonics $h = 7$ to
$11$ of Fig.~\ref{fig:spectrum}; \iaopp{} stays below
$\pm$\SI{0.85}{\ampere} with a visibly more uniform trace. The effect
arises almost entirely from the placement of the edges and the choice of
the step sizes: as laid out in \eqref{eq:vreq}, the pattern aims at
the predistorted voltage shape that produces the ideal current on the
real plant. Exploitation of the fundamental tolerance band contributes
only marginally at this budget (below 0.01 percentage points); it
becomes visible as a partial compensation of the common
fundamental-response deviation discussed in
Section~\ref{ssec:chosen}.

At $n = 48$ the two methods are separated by \SI{0.607}{\percent}
versus \SI{0.521}{\percent}, a relative advantage of
\SI{14}{\percent}; \iaopp{} undercuts the floor of the
voltage-oriented methods by close to \SI{8}{\percent} and is the only
method in the field to do so. The gap widens at small budgets
($n = 12$: \SI{2.00}{\percent} versus \SI{1.07}{\percent}; $n = 8$:
\SI{5.00}{\percent} versus \SI{1.99}{\percent}), because fewer edges
make their placement more consequential. The value of
plant modeling is thereby isolated: with optimizer,
budget, and evaluation identical, the model in the objective alone
accounts for the prediction gap (factor 10), the broken floor, and the
efficiency of small budgets (factor 2 to 2.5).

\subsection{Benchmark}
\label{ssec:benchmark}

\subsubsection{The THD Plane: Current Quality Versus Bus Load}

\begin{figure}[!t]
  \centering
  \paperfig{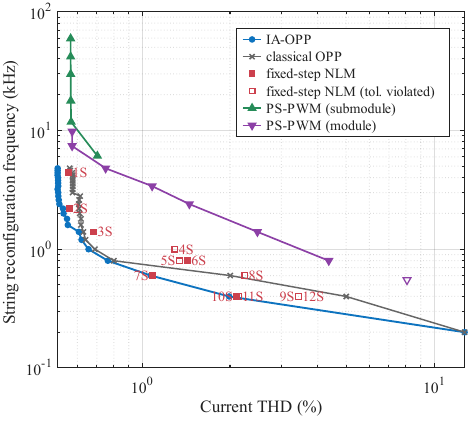}
  \caption{THD plane at the nominal operating point: string
    reconfiguration frequency versus current THD (both logarithmic) for
    \iaopp{}, classical OPP, fixed-step NLM, and PS-PWM at submodule and
    module level; open markers violate the fundamental tolerance.}
  \label{fig:thdplane}
\end{figure}

Fig.~\ref{fig:thdplane} plots, for every method, the string
reconfiguration frequency (the bus component of the dual
metric) against the current THD; both axes are logarithmic. Each method
family appears as a curve
over its respective effort parameter: \iaopp{} over the budget $n = 4$
to $96$, the classical OPP baseline over the same budgets, NLM over the
step size $S$, and the two PS-PWM variants over the frequency ratio
$m_f$. Open markers denote configurations violating the fundamental
tolerance.

The \iaopp{} curve forms the
Pareto front: at every bus load it attains the lowest THD, and for every
demanded THD the lowest bus load. 1S-NLM reaches \SI{0.561}{\percent} at
\SI{4.4}{\kilo\hertz}; \iaopp{} undercuts this with
\SI{0.521}{\percent} at \SI{2.4}{\kilo\hertz}, better quality at
just under half the bus load; 2S-NLM at comparable bus load
(\SI{2.2}{\kilo\hertz}) lies 0.04 percentage points above. Coarser
fixed-step staircases ($S \ge 4$) already fail feasibility: with fixed
large steps, the fundamental amplitude can no longer be met within
\SI{1}{\percent} at this operating point, an immediate argument for
optimized step sizes. Submodule-level PS-PWM illustrates the floor of
Section~\ref{ssec:plantvalue}: from $m_f = 2$ its curve
is vertical; a fivefold increase of the switching effort to beyond
\SI{60}{\kilo\hertz} buys no further current quality. Module-level
PS-PWM, the polarity-stage reference, runs to the right of the field
throughout. The classical OPP accompanies the \iaopp{} curve
at a constant offset and merges into the floor with growing budget;
this is the gap of Section~\ref{ssec:plantvalue} across all budgets.
The Pareto front has a knee around $n = 32$ to
$48$; above it the gain flattens, consistent with the selected
$n = 48$.

\subsubsection{Operating-Point Sweep}

\begin{figure}[!t]
  \centering
  \paperfig{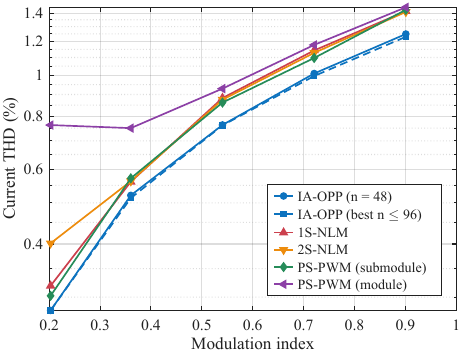}
  \caption{Operating-point sweep: current THD versus modulation index
    for \iaopp{} (fixed budget $n = 48$ and best budget $n \le 96$),
    fixed-step NLM, and PS-PWM at submodule and module level.}
  \label{fig:oppoints}
\end{figure}

A lookup-table method must hold up across the modulation range.
Fig.~\ref{fig:oppoints} shows the current THD for modulation indices
$m_\mathrm{a} = 0.2$ to $0.9$ (\SIrange{45}{200}{\volt}): \iaopp{} is
re-optimized per operating point, following the lookup-table
concept, with the budget held fixed at $n = 48$; additionally plotted
is the best budget $n \le 96$. The reference methods scale by their own
logic: the NLM staircases follow the reference (their event count grows
with the modulation index), the PS-PWM variants keep their $m_f$.

Three findings. First, \iaopp{} with fixed budget is the best method at
every operating point. At the lower end ($m_\mathrm{a} = 0.2$) it
attains
\SI{0.279}{\percent} against \SI{0.319}{\percent} for 1S-NLM, which
there uses the same budget of 48 events: the advantage stems purely from
the better placement of equally many edges. At the upper end
($m_\mathrm{a} = 0.9$) the figures are \SI{1.25}{\percent} against
\SI{1.41}{\percent}, with 1S-NLM requiring 216 events there, 4.5 times
the bus load. Second, the fixed budget is practically optimal: the curve
of the best budget $n \le 96$ lies almost on top of the $n = 48$ curve
(1 to \SI{2}{\percent} better in relative terms; at the lower end,
$n = 48$ is exactly the best budget). A single budget carries across
the entire modulation range, which simplifies the lookup table
accordingly. One property of the exact budget binding deserves
mention: a \emph{forced} large budget at small modulation degenerates
(at $m_\mathrm{a} = 0.2$, $n = 96$ forces needle pulses and
\SI{7.2}{\percent} THD); the budget selection must admit small budgets,
as the selection rule and the target-THD mode do. Third, the sweep
confirms the layer assignment: module-level PS-PWM, which switches
whole modules through the polarity stage, is the weakest method
throughout and falls behind at small modulation (flat at
\SI{0.763}{\percent} against \SI{0.279}{\percent} for \iaopp{}, factor
2.7). Its carrier ripple does not scale with the modulation index, and
it lacks fine submodule-level actuation steps.

\subsection{Robustness}
\label{ssec:robustness}

A method that predistorts its patterns on a plant model is committed to
that model; the constant-resistance experiment of
Section~\ref{ssec:plantvalue} showed how a predistortion turns into a
disadvantage when the modeled property vanishes entirely. For
lookup-table operation it must therefore be examined how parameter
deviations of realistic size play out. To this end, the patterns
designed at the nominal point (the \iaopp{} solution, $n = 48$, and all
reference methods) are evaluated unchanged on perturbed plants: load
inductance $L$, load resistance $R$, cell internal resistance
$r_\mathrm{bat}$, and cell voltage each vary by $\pm$\SI{30}{\percent},
first individually, then in all 16 simultaneous combinations of the
extreme values. Rating again uses the identical closed-form evaluation.

In the single-parameter sweeps, \iaopp{} remains the best method at
every point of every parameter; its worst value is
\SI{0.661}{\percent}, while the best baseline at the same point does not
drop below \SI{0.703}{\percent}. The predistortion thus degrades
gracefully: a \SI{30}{\percent} model deviation narrows the advantage
but nowhere reverses it. The cell voltage is a special case: it scales
the entire staircase voltage and with it the current proportionally, so
the THD as a ratio remains exactly invariant. Affected is solely the
fundamental amplitude, whose tracking in lookup-table operation falls to
the amplitude control and is not a pattern-shape question.

The 16 simultaneous corner points are the harder test because the
deviations accumulate. \iaopp{} is the best method in 12 of the 16
corners; its worst case over all corners is \SI{0.945}{\percent} against
\SI{0.993}{\percent} for the best baseline. Even under maximum
simultaneous perturbation, the nominally designed solution remains the
most reliable choice. The four exceptions are
the benign corners with increased inductance and
decreased cell resistance: more filtering, weaker resistance
modulation. There, precisely the effect that \iaopp{} exploits shrinks;
all methods lie far below their nominal THD, and 1S-NLM edges ahead by
at most 0.017 percentage points. \iaopp{} loses ground only where the
problem it solves has become smaller, and then only marginally; the
constant-resistance experiment of Section~\ref{ssec:plantvalue} marks
the extreme end of this spectrum.

For operation this means: within a parameter uncertainty of
$\pm$\SI{30}{\percent}, the lookup-table patterns require neither
adaptation nor re-identification; the advantage over the reference
methods is retained in nearly all constellations, and where it is not,
the deficit is orders of magnitude smaller than the gain at the nominal
point. Systematic parameter drift, e.g., through aging, can
be absorbed by occasional offline recomputation of the lookup table
without touching running operation.

\section{Three-Phase Operation}
\label{sec:threephase}

The results so far hold for the individual string and thus directly for
single-phase systems, four-wire configurations with neutral conductor,
and delta connection. In a three-wire star connection an additional
question arises: harmonics of triplen order are common-mode quantities
there and drive no line current, and it is precisely the dominant third
harmonic of the resistance-modulation floor that falls to the zero
sequence. The following analysis is of zeroth order: the triplen orders
are removed from the exactly computed single-string spectra (patterns
shifted by $T/3$ per phase, feedback of the phase coupling neglected);
the triplen voltages persist as common mode and shift the star point,
but drive no line current.

The numbers confirm this: the floor is, to 88 to
\SI{98}{\percent} of its distortion energy, a triplen phenomenon. Of
the \SI{0.565}{\percent} floor, only about \SI{0.09}{\percent} remains
in the three-wire star. For the patterns optimized in the single-phase
setting, the advantage shrinks accordingly: \iaopp{} stays ahead of
classical OPP at equal budget (\SI{0.109}{\percent} against
\SI{0.177}{\percent} non-triplen THD at $n = 48$) but is narrowly
undercut by submodule-level PS-PWM (\SI{0.082}{\percent} at
\SI{11.8}{\kilo\hertz}) and by classical OPP at double budget
(\SI{0.089}{\percent} at $n = 96$).

This computation, however, measures the wrong task: a pattern that
spends budget on compensating a harmonic that the grid filters anyway
optimizes past the goal. The grid topology belongs in the objective,
like the impedance. With triplen-masked THD in the objective
(orders $3, 9, 15, \ldots$ excluded from \eqref{eq:thd} during
refinement and selection), the three-phase-aware \iaopp{} restores the
order of the field: at $n = 48$ it attains \SI{0.076}{\percent} and
thereby undercuts every method in the field, including submodule-level
PS-PWM (\SI{0.082}{\percent}) at one fifth of its bus load; even
$n = 40$ (\SI{2.0}{\kilo\hertz}) beats the PWM. With free
budget selection, the selection rule picks $n = 92$ at
\SI{0.051}{\percent}, against \SI{0.063}{\percent} at $n = 96$ for the
best feasible linear-model counterpart. The optimizer
exploits the mask: evaluated single-phase, the same pattern
exhibits \SI{1.44}{\percent}; the distortion is shifted
into the currentless triplen orders. The patterns are therefore specific
to the grid topology: the plant and the grid a pattern
will see belong among its design assumptions.

The fair yardstick is again the single-variable experiment: the same
mask, but a linear plant model in the objective.
Table~\ref{tab:threephase} shows the result: the impedance advantage
survives the three-phase case and even grows in relative terms, from
\SI{14}{\percent} to about \SI{21}{\percent} at $n = 48$ and
consistently 15 to \SI{30}{\percent} over the operating range
$n \ge 24$. The reason: the common triplen floor
disappears, and the remaining orders $5, 7, 11, \ldots$ are
those on which impedance-aware edge placement acts. The
prediction gap of the linear model also persists in three-phase
operation (promised \SI{0.031}{\percent}, delivered
\SI{0.098}{\percent} at $n = 48$, a factor of 3).\footnote{At very
small budgets ($n \le 20$) both three-phase-aware runs scatter due to
local optima of the timing refinement; reliable statements there require
multi-start optimization. The operating range $n \ge 24$ considered here
is unaffected; cf.\ the multi-start probe at $n = 48$ in
Section~\ref{ssec:twostage}, which scatters by only 0.004 percentage
points.}

\begin{table}[!t]
  \caption{Three-phase-aware optimization (triplen-masked THD in the
    objective): non-triplen THD on the real plant for \iaopp{} and for
    classical OPP with the identical mask.}
  \label{tab:threephase}
  \centering
  \renewcommand{\arraystretch}{1.15}
  \setlength{\tabcolsep}{5pt}
  \small
  \begin{tabular}{lccc}
    \toprule
    Budget $n$ & Classical OPP & \iaopp{} & Advantage \\
    \midrule
    48 & \SI{0.098}{\percent} & \textbf{\SI{0.076}{\percent}} & \SI{22}{\percent} \\
    64 & \SI{0.079}{\percent} & \textbf{\SI{0.062}{\percent}} & \SI{21}{\percent} \\
    96 & \SI{0.063}{\percent} & \textbf{\SI{0.050}{\percent}} & \SI{21}{\percent} \\
    \bottomrule
  \end{tabular}
\end{table}

\section{Discussion and Conclusion}
\label{sec:conclusion}

\subsection{Summary of Findings}

This paper has substantiated its thesis: the class of optimized pulse
patterns fits the reconfigurable battery system, but only the inclusion
of the system's structural properties in the optimization unlocks its
potential. The value of plant modeling was not asserted but measured in
a single-variable experiment: against classically formulated OPP at
identical budget stands a quality advantage of \SI{14}{\percent} at the
nominal point up to a factor of 2.5 at small budgets; the prediction of
the linear model misses reality by up to a factor of 10; and the THD
floor of the voltage-oriented methods, traced back completely to the
level-synchronous resistance modulation, is undercut by \iaopp{} as the
only method in the field. The event budget with \rev{optimized step
sizes} places
\iaopp{} on the Pareto front of the benchmark (the lowest bus load for
every demanded current quality), the robustness
analysis shows that the predistortion degrades gracefully within
$\pm$\SI{30}{\percent} parameter uncertainty, and the three-phase
extension demonstrates that the impedance advantage survives, and even
grows, once the grid topology joins the impedance in the objective.

\subsection{Limitations}

Three limitations accompany the results. First, this is a simulation
study: all quantities derive from an exact analytical evaluation of the
switched RL plant, but no experimental validation is available;
measurements on a hardware demonstrator remain future work. Within the
simulation scope, the RL surrogate represents the motor-type load only
stationarily; back-EMF and torque formation are outside the model. The
plant model is idealized in two further respects: the cell impedance is
represented by a purely ohmic internal resistance, whereas real cell
impedance is frequency- and state-of-charge-dependent, and switching
transitions are ideal (no dead times or commutation transients). The
robustness analysis addresses the first simplification in part---the
advantage survives $\pm$\SI{30}{\percent} resistance mismatch---but an
impedance-spectrum-aware objective, immediately formulable with the same
machinery, remains an extension for future work.
Second, \iaopp{} is an offline method for stationary operating points
from a feedforward perspective: the transition between lookup-table
patterns upon operating-point changes, a classical problem area of the
OPP family, was not examined; the transition strategies established
there \cite{holtz2007} should carry over, but the demonstration is
pending. A closed current-control loop could compensate part of the
plant mismatch online, at the cost of bandwidth and without the
deterministic event guarantees that motivate the approach here. Third,
the method optimizes the current THD; the losses are computed exactly
but do not enter the objective. A multi-criteria extension is
immediately formulable with the same evaluation machinery. In the same
category belongs state-of-charge balancing: the optimized level sequence
determines \emph{how many} cells are inserted at any time, not
\emph{which}; the assignment of specific submodules remains available
as a degree of freedom for balancing rotation, whose interaction with
the switching metric was not examined here. The three-phase analysis,
finally, is of zeroth order; an exact coupled three-phase evaluation
with free star point is future work.

\subsection{Transferability}

The formulation is not tied to the reference system. The plant enters
the objective exclusively through the mapping from voltage level to path
resistance \eqref{eq:pathres}; any system class whose source impedance
depends on the switching state merely supplies a different such mapping.
This includes reconfigurable systems without a polarity stage: the
modulation operates almost entirely on the submodule layer anyway, the
H-bridges steering only the polarity; for a unipolar system the polarity
mirroring is dropped while the quarter-wave structure and the
optimization core remain unchanged (a demonstrating computation is
pending). The resistance mapping is likewise extensible to additional
configuration degrees of freedom, e.g., parallel module groups whose
path resistance does not grow monotonically with the level; the dynamic
program does not require monotonicity of the mapping. The three-phase
analysis has already exercised this modularity: grid topology and plant
impedance are two independent configurations of the same objective, and
only their combination defines the correct optimization problem in each
case.

In summary, impedance-aware optimized pulse patterns establish
themselves as the natural stationary modulation strategy of
reconfigurable battery systems: the method inherits the maturity and
lookup-table principle of the OPP family, takes the structural
properties of the system (configuration-dependent impedance,
event-based cost structure, and grid topology) into the optimization,
and achieves a current quality that remains out of
reach for voltage-oriented methods on this plant class, in the single
string as in the three-phase arrangement.

\section*{Acknowledgment}
This research out of the project MORE (Munich Mobility Research Campus)
is funded by dtec.bw---Digitalization and Technology Research Center of
the Bundeswehr, which we gratefully acknowledge. dtec.bw is funded by
the European Union---NextGenerationEU.

The authors acknowledge the use of a large language model (Claude,
Anthropic) to assist in drafting and editing the text of this
manuscript \rev{and in implementing the accompanying MATLAB code}. All
methods, results, and conclusions were
developed and verified by the authors, who take full responsibility for
the content of this article \rev{and of the published code; the
implementation is verified by its regression-test suite and by the
independent cross-validation of Section~\ref{ssec:closedform}}.

\rev{The MATLAB implementation and the result data are openly available
at \url{https://github.com/pelican069/iaopp-rbs} and permanently
archived at \url{https://doi.org/10.5281/zenodo.24338849}.}

\bibliographystyle{IEEEtran}
\bibliography{references}

\end{document}